\documentclass[aps,nofootinbib,11pt,preprintnumbers]{revtex4}

\usepackage{bm}
\usepackage{latexsym}
\usepackage{amsfonts}
\usepackage{amssymb}
\usepackage{amsmath}
\usepackage{color}

\begin{document}

\preprint{YITP-16-111}

\title{Hamiltonian Analysis of an On-shell $U(1)$ Gauge Field Theory}
\author{Chunshan Lin}
\email{chunshan.lin@yukawa.kyoto-u.ac.jp}
\author{Misao Sasaki}
\email{misao@yukawa.kyoto-u.ac.jp}
\affiliation{Center for Gravitational Physics,
Yukawa Institute for Theoretical Physics, Kyoto University}

\date{\today}

\begin{abstract}
\begin{center}
~\\
\Large{Abstract}
\end{center}
We perform the Hamiltonian analysis of an on-shell $U(1)$ gauge field theory, 
in which the action is not invariant under local $U(1)$ transformations
but recovers the invariance when the equations of motion are imposed. 
We firstly apply Dirac's method of Hamiltonian analysis. 
We find one first-class constraint and two second-class constraints
 in the vector sector. It implies the photons have only two polarisations, 
at least at the classical level, although the standard $U(1)$ symmetry is 
explicitly broken. The reduced Hamiltonian is bounded from below 
and the on-shell $U(1)$ gauge field theory is free from ghosts at
the classical level.
\end{abstract}

\maketitle

\newpage
\section{Introduction}
In quantum field theory, the local $U(1)$ gauge symmetry is the most established
 symmetry underlying the interaction between matter and
the eletromagnetic field.
 There are several profound physical implications including the current 
conservation and the absence of longitudinal mode of photon and so on. 
At the quantum level, the Ward-Takahashi identity protects the masslessness 
of photons against quantum loop corrections and thus a photon can find no rest.   

At high energy scales, the break-down of perturbation theory due
to the Landau pole singularity implies that our QED is 
just a low energy effective theory and it should be integrated into
 a larger symmetry group of $SU(2)\otimes U(1)$ \cite{SM}.
 On the other hand, the theory exhibits less symmetry at ultra low energy scale.
 For example, as known in condense matter physics, pairs of electrons may
be driven into boson condensates described by a $U(1)$-charged complex 
scalar field at low energy scales. 
Below a critical temperature, the scalar field develops a non-trivial 
VEV and the $U(1)$ symmetry is spontaneously broken. 
As a consequence, some materials exhibit superconductivity \cite{GL}. 
A theory with an explicitly broken $U(1)$ symmetry was proposed by Proca 
in his work on the massive spin-1 boson field~\cite{Proca:1900nv}. 
In addition to the 2 transverse degrees of freedom, the massive spin-1 
particle has the longitudinal mode, which decouples
from the scattering process in the massless limit. 

It is then intriguing to ask whether there are some other symmetry breaking 
mechanisms of which the broken phase is less symmetric than QED but 
more symmetric than the Proca theory? Inspired by this question, 
we come up with an idea of a ``weakly broken" gauge symmetry. 
By ``weakly broken" we mean that the action of our theory is invariant 
under local $U(1)$ transformations only when the equations of motion are 
imposed. We call this theory an on-shell $U(1)$ gauge field theory.
A very natural and direct question  is that how many degrees of freedom 
are there in this type of gauge field theory?  
This question should be addressed at both classical and quantum levels. 
In this paper, we consider an interesting realization of such a theory
and perform the Hamiltonian analysis at classical level.
 The computation of quantum corrections is deferred to future work.

This paper is organised as follows: In Sec.~\ref{onshell}, we will briefly 
introduce the idea of on-shell gauge symmetry. In Sec.~\ref{dirac}, 
we apply Dirac's method of Hamiltonian analysis and show that the number of 
degrees of freedom remains the same as the usual scalar plus $U(1)$ gauge theory.
 We derive the reduced Hamiltonian in Sec.~\ref{FJ}, and show that the Hamiltonian is bounded from below. Conclusions and further 
discussions are given in Sec.~\ref{cd}.

\section{The on-shell $U(1)$ gauge symmetry}\label{onshell}
In our theory, there is a scalar field which breaks the local $U(1)$ 
symmetry explicitly but ``weakly". The Lagrangian density is written as 
\begin{eqnarray}\label{modQED}
\mathcal{L}=&&
-\frac{1}{4}g^2(\phi)F_{\mu\nu}F^{\mu\nu}
- e_0f(\phi)A_{\mu}\bar{\psi}\gamma^{\mu}\psi
+i\bar{\psi}\gamma^{\mu}\partial_\mu\psi-m\bar{\psi}\psi
-\frac{1}{2}g^{\mu\nu}\partial_\mu\phi\partial_\nu\phi-V(\phi)\,,
\end{eqnarray}
where $g(\phi)$ and $f(\phi)$ are two generic and dimensionless function 
of a scalar field $\phi$. For simplicity, we only consider a single
charged Dirac fermion, say electron, as a representative of matter, and
ignore the gravitational degrees of freedom since they are irrelevant to 
the current issue. 
The local $U(1)$ symmetry of QED is broken due to the abnormal gauge 
coupling $e_0f(\phi)A_{\mu}\bar{\psi}\gamma^{\mu}\psi$. 

This model,
with the special choice $f(\phi)=g(\phi)$, was 
initially proposed to explain the ubiquitous presence of cosmic magnetic 
fields \cite{Domenech:2015zzi}. The pre-factor $g(\phi)^2$ in front of 
$F_{\mu\nu}F^{\mu\nu}$ increase exponentially and compensate the redshift 
factor of the magnetic field due to the expansion during inflation, 
while the strong coupling between the gauge field and the fermion 
can be avoided by setting $f(\phi)=g(\phi)$ in 
$e_0f(\phi)A_{\mu}\bar{\psi}\gamma^{\mu}\psi$.  
Related to our model, a similar $U(1)$ symmetry breaking model
was discussed earlier by Dvali~\cite{Dvali:2004tma}, which was motived to 
solve the gauge hierarchy problem, i.e. why the electroweak scale is so
 much lower than the Planck scale.  In the model of  Ref.~\cite{Dvali:2004tma},
 a three-form field interacts with 2-branes, where the charge of the latter 
is a function of a scalar field and thus the local gauge symmetry 
is also explicitly broken. 

To understand the internal symmetry of this theory, let us
firstly derive the equations of motion of the vector field,
\begin{eqnarray}
\partial_\mu\left(g^2F^{\mu\nu}\right)-e_0f\bar{\psi}\gamma^\nu\psi=0.
\end{eqnarray}
Taking the divergence of the above, by noting the identity
$\partial_\mu\partial_\nu F^{\mu\nu}\equiv0$, we must have 
\begin{eqnarray}\label{current1}
\partial_\mu\left(e_0f\bar{\psi}\gamma^\mu\psi\right)=0.
\end{eqnarray}
On the other hand, the global $U(1)$ symmetry in the fermion sector 
implies the existence of a conserved Noether current, 
\begin{eqnarray}\label{current2}
\partial_\mu\left(\bar{\psi}\gamma^{\mu}\psi\right)=0.
\end{eqnarray}
Combining eqs. (\ref{current1}) and (\ref{current2}), we obain the following 
``constraint'' equation:
\begin{eqnarray}\label{current3}
\bar{\psi}\gamma^\mu\psi\partial_\mu f=0.
\end{eqnarray}

Under the local $U(1)$ gauge transformation,
\begin{eqnarray}
A_\mu\to A_\mu+\partial_\mu\chi,\qquad\psi\to e^{-ie_0f\chi}\psi,
\end{eqnarray}
the variation of the action reads
\begin{eqnarray}
\delta_\chi S=e_0\int d^4x\chi
\left(\bar{\psi}\gamma^\mu\psi\partial_\mu f\right)+\text{total derivatives}.
\end{eqnarray}
Where $\delta_\chi$ denotes the variation $\delta_\chi A_\mu=\partial_\mu\chi$. 
With Eq.~(\ref{current3}) imposed, the variation of the action vanishes. 
Therefore, our action is invariant under local $U(1)$ gauge transformations 
only when the equations of motion are imposed. 

Here two comments are in order. One may be confused by our statement  about the on-shell gauge symmetry. One may think that with help of equation of motion, any actions are invariant under the transformation of fields. However, that is only the case when we perform the linear and infinitesimal transformation of the field. As a gauge invariant theory, the action must be invariant at fully non-linear level. In our theory, we do need equations of motion to prove that our theory is invariant at fully non-linear level.  On the other hand,  Since it looks like eq.~(\ref{current3}) 
contrains the system too much, one might worry if it would render 
the dynamics trivial. In the following, we show this is not the case. 
We show that the system has just the necessary and sufficient number of
dynamical degrees of freedom.
In fact, since eq.~(\ref{current3}) guarantees
the on-shell $U(1)$ symmetry as mentioned above, it could be regarded 
as a constraint that kills the longitudinal component of the vector field 
which would be present if $U(1)$ were totaly broken.

\section{Dirac's method of Hamiltonian Analysis}\label{dirac}
In this section, we apply Dirac's method of Hamiltonian analysis~\cite{dirac}
to our theory~(\ref{modQED}).
The conjugate momenta read
\begin{eqnarray}\label{modHam}
&&\pi_\phi=\frac{\delta\mathcal{L}}{\delta \dot{\phi}}
=\dot{\phi},
\qquad\pi_i=\frac{\delta\mathcal{L}}{\delta\dot{A}_i}
=g^2\left(\dot{A}_i-\partial_iA_0\right),
\qquad\pi_0=\frac{\delta\mathcal{L}}{\delta \dot{A}_0}=0,
\nonumber\\
&&\pi_\psi=\frac{\delta\mathcal{L}}{\delta\dot{\psi}}=i\psi^\dagger,
\qquad\qquad\pi_{\psi^\dagger}=\frac{\delta\mathcal{L}}{\delta\dot{\psi^\dagger}}=0.
\end{eqnarray}
The Hamiltonian density is thus given by 
\begin{eqnarray}
\mathcal{H}_0=&&\frac{1}{2}\pi_\phi^2
+\frac{1}{2}\partial_i\phi\partial_i\phi+V(\phi)
+\frac{1}{2}\pi_i\pi_i+g^2\left(\partial_iA_j-\partial_iA_j\partial_jA_i\right)
-A_0\partial_i\pi_i\nonumber\\
&&+e_0f(\phi)A_\mu\bar{\psi}\gamma^\mu\psi-i\bar{\psi}\gamma^i\partial_i\psi
+m\bar{\psi}\psi+\lambda_0\pi_0+\lambda_\psi\left(\pi_\psi-i\psi^\dagger\right)
+\pi_{\psi^\dagger}\lambda_{\psi^\dagger},
\end{eqnarray}
with the 3 primary constraints,
\begin{eqnarray}
\Phi_1=\pi_0\approx0,\qquad
\Phi_2=\pi_\psi-i\psi^\dagger\approx0,\qquad
\Phi_3=\pi_{\psi^\dagger}\approx0\,,
\end{eqnarray}
where $\approx0$ denotes the equality when the constraints are satisfied.
To be consistent, we demand the Poisson brackets of these 3 primary 
constraints with the Hamiltonian to vanish, i.e. 
$\{\Phi_i,H\}_{P.B.}\approx0$. 

Let us first look at the fermion sector, 
\begin{eqnarray}
\{\Phi_2,H_0\}_{P.B.}
&=&-i\lambda_{\psi^{\dagger}}-\left[e_0g(\phi)A_\mu\bar{\psi}\gamma^\mu
+i\partial_i\left(\bar{\psi}\gamma^i\right)+m\bar{\psi}\right],\\
\{\Phi_3,H_0\}_{P.B.}
&=&i\lambda_{\psi}-\left[e_0g(\phi)A_\mu\gamma^0\gamma^\mu\psi
-i\gamma^0\gamma^i\partial_i\psi+m\gamma^0\psi\right].
\end{eqnarray}
These 2 Poisson brackets fix the coefficients $\lambda_{\psi}$ 
and $\lambda_{\psi^{\dagger}}$ and therefore they do not give any new
 constraints. Note that these 2 primary constraints, as well as the 
determined values of $\lambda_\psi$ and $\lambda_{\psi^\dagger}$ are 
essential to obtain the correct equations of motion for the fermion.
    
Now let us compute the Poisson brackets of the first primary 
constraint with the Hamiltonian. For the primary constraint $\pi_0\approx0$, 
we have consistency condition,
\begin{eqnarray}\label{phi4}
\{\Phi_1,H_0\}_{P.B.}
=e_0f(\phi)\psi^\dagger\psi-\partial_i\pi_i\equiv\Phi_4\approx 0.
\end{eqnarray}
Thus the consistency condition gives us a secondary constraint and we name 
it $\Phi_4$.
Then we plug it into the Hamiltonian and treat it as the same footing 
as the primary constraints,
\begin{eqnarray}
\mathcal{H}_1\equiv\mathcal{H}_0+\lambda_4\Phi_4.
\end{eqnarray}
The consistency condition requires that the Poisson bracket of this new 
secondary constraint $\Phi_4$ with the Hamiltonian must also vanish, 
and it further gives one more secondary constraint,
\begin{eqnarray}\label{phi5}
\{\Phi_4,H_1\}_{P.B.}
=e_0f'\left[\psi^\dagger\psi\cdot\pi_{\phi}
+\bar{\psi}\gamma^i\psi \partial_i\phi\right]\equiv\Phi_5\approx0.
\end{eqnarray}
Again, we plug the new secondary constraint into the Hamiltonian, 
and treat it as the same footing as the primary constraints.
Then we obtain the total Hamiltonian
\begin{eqnarray}
\mathcal{H}_T\equiv \mathcal{H}_1+\lambda_5\Phi_5.
\end{eqnarray}
The consistency condition 
\begin{eqnarray}
\{\Phi_i,H_T\}_{P.B}\approx0,\qquad\text{where}\qquad i=1,2,3,4,5,
\end{eqnarray}
fixes the coefficients $\lambda_\psi$, $\lambda_{\psi\dagger}$, $\lambda_4$ 
and $\lambda_5$ and no new secondary constraints are 
generated\footnote{We need to be careful here 
$\{\Phi_5,H\}_{P.B}\supset\{\Phi_5,\int \lambda_5\Phi_5 d^3x\}+...
=-\lambda_5e_0g'\bar{\psi}\gamma^i\psi
\partial_i\left(e_0g'\psi^\dagger\psi\right)+...$
 and thus the consistency condition of $\Phi_5\approx0$ fixes $\lambda_5$. }.
 We also find that 
\begin{eqnarray}
\{\Phi_1,\Phi_i\}\approx0,\qquad\text{where}\qquad i=1,2,3,4,5.
\end{eqnarray}
Therefore, $\Phi_1\approx0$ is the first-class constraint. 
On the other hand,
\begin{eqnarray}
&&\{\Phi_m,\Phi_n\}\simeq\nonumber\\
&&\left(\begin{array}{cccc}
0 & -i & e_{0}f\psi^{\dagger} & -\left(e_{0}f'\psi^{\dagger}\pi_{\phi}
+e_{0}\bar{\psi}\gamma^{i}\partial_{i}f\right)\\
i & 0 & e_{0}f\psi & -\left(e_{0}f\psi\pi_{\phi}
+e_{0}\gamma^{0}\gamma^{i}\psi\partial_{i}f\right)\\
-e_{0}f\psi^{\dagger} & -e_{0}f\psi & 0 & 
-\left(e_{0}f'\psi^{\dagger}\psi\right)^{2}\\
\left(e_{0}f'\psi^{\dagger}\pi_{\phi}
+e_{0}\bar{\psi}\gamma^{i}\partial_{i}f\right) 
& \left(e_{0}f\psi\pi_{\phi}+e_{0}\gamma^{0}\gamma^{i}\psi\partial_{i}f\right)
 & \left(e_{0}f'\psi^{\dagger}\psi\right)^{2} & 0
\end{array}\right),\nonumber\\
&&\text{where}\qquad m,n=2,3,4,5.
\end{eqnarray}
The determinant of the above matrix is non-vanishing,
\begin{eqnarray}
\det\{\Phi_m,\Phi_n\}=-\left(e_{0}f'\psi^{\dagger}\psi\right)^{4}\neq0,
\end{eqnarray}
and therefore all of them are second-class. 

Now we are ready to count the number of the degrees of freedom. 
Firstly, the number of degrees of fermion is the same as the one in 
standard QED,  $\Phi_2\approx0$ and $\Phi_3\approx0$ are second class, 
they fix the 1-to-1 correspondence between canonical variables and their 
conjugate momenta. Therefore the fermion has 4 states, 2 for electron and 
another 2 for positron, which is the same as the standard QED. 
In the vector field sector, the Hamiltonian contains 8 canonical conjugate pairs.
 We have 1 first-class constraint and 2 second-class constraints.
Together they kill another 4 degrees in phase space, thus the number of degrees 
of freedom in the phase space of the electromagnetic field is 4. 
The corresponding number of physical degrees of freedom is thus 2. 
Adding the canonical conjugate pair of the scalar $\phi$, we conclude 
that  the  number of degrees of freedom in our theory is  the same as 
the usual scalar plus $U(1)$ gauge theory. Namely, no longitudinal mode is found in the photon of our theory. 

The Hamiltonian analysis has helped us to understand the physical meaning of on-shell U(1) symmetry. The on-shell U(1) symmetry gives us  two second class constraint equations (or in other words, conservation laws), they eliminate the longitudinal mode of photon.


\section{Reduced Hamiltonian and Ghost Freeness}\label{FJ}
We have proved that our theory has correct number of degrees of freedom in the previous sections. The validity of the theory, at least at classical level, requires the Hamiltonian must be bounded from below. 
Since we are dealing with system with constraints, we have to integrate out all constraints, and derive the reduced Hamiltonian. In this approach, one rewrites the Lagragian 
in the first order form, and reduces it by plugging all the solutions
of the constraint equations into it, 
irrespective of whether they are first-class or second-class.

Again, we start from the Lagrangian~(\ref{modQED}). 
The conjugate momenta read
\begin{eqnarray}\label{momentum1}
\pi_\phi=\dot{\phi}\,,\quad
\pi_i=g^2\left(\phi\right)\left(\dot{A}_i-\partial_iA_0\right)\,,
\quad\pi_0=0\,,
\quad
\pi_\psi=i\psi^{\dagger}\,,
\quad\pi_{\psi^\dagger}=0\,.
\end{eqnarray}
As before the last 3 momenta form 3 primary constrains.
Plugging them into the Lagrangian, the Hamiltonian reads
\begin{eqnarray}
H=&&\frac{1}{2}\pi_\phi\pi_\phi
+\frac{1}{2}\partial_i\phi\partial_i\phi+V(\phi)
+\frac{1}{2g^2}\pi_i^2
+\frac{1}{2}g^2\partial_iA_j\partial_iA_j
-\frac{1}{2}g^2\partial_iA_j\partial_jA_i
+e_0f(\phi)A_i\bar{\psi}\gamma^i\psi
\nonumber\\
&&-i\bar{\psi}\gamma^i\partial_i\psi+m\bar{\psi}\psi
+A_0\left[-\partial_i\pi_i+e_0f(\phi)\psi^{\dagger}\psi\right]
\nonumber\\
\equiv
&&H_1+A_0\left[-\partial_i\pi_i+e_0f(\phi)\psi^{\dagger}\psi\right]\,,
\label{HamFJ}
\end{eqnarray}
where and in what follows,
$\bar\psi=\psi^\dag\gamma^0=-i\pi_\psi\gamma^0$ is understood.
The Lagrangian in the first order form is given by
\begin{eqnarray}\label{la0}
\mathcal{L}=\pi_\phi\dot{\phi}+\pi_i\dot{A}_i+\pi_\psi\dot{\psi}-H_1
+A_0\left[\partial_i\pi_i-e_0f(\phi)\psi^{\dagger}\psi\right]\,.
\end{eqnarray}

As clear from the above, there appears a secondary constraint
with $A_0$ as a Lagrange multiplier,
\begin{eqnarray}\label{a0cons}
\partial_i\pi_i=e_0f(\phi)\psi^{\dagger}\psi\,.
\end{eqnarray}
To solve this constraint we decompose the gauge field and its conjugate 
momentum into the transverse and longitudinal parts, 
\begin{eqnarray}
A_i=A_i^{T}+\partial_i\chi,\hspace{10mm}\pi_i=\pi_i^T+\partial_i\pi,
\end{eqnarray}
where $\partial^iA_i^T=\partial^i\pi_i^T=0.$ 
Then the constraint~(\ref{a0cons}) can be rewritten as 
\begin{eqnarray}
\pi=\Delta^{-2}\left(e_0f\psi^\dagger\psi\right),
\label{pidef}
\end{eqnarray}
where $\Delta=\partial^i\partial_i$ 
and $\Delta^{-2}$ is a non-local operator defined in the way such 
that $\Delta\Delta^{-2}Q=Q$ with an appropriate boundary condition.
The Lagrangian is now rewritten as 
\begin{eqnarray}
\mathcal{L}_1
=\pi_\phi\dot{\phi}+\pi_i^T\dot{A}_i^T+\pi_\psi\dot{\psi}
-H_{red}+\chi e_0\left[-\partial_t \left(f\psi^\dag\psi\right)
+\partial_i\left(f\bar\psi\gamma^i\psi\right)\right],
\label{Lag1}
\end{eqnarray}
where we have employed integration by parts appropriately and
used Eq.~(\ref{pidef}).
The $H_{\text red}$ is the reduced Hamiltonian, 
\begin{eqnarray}\label{tildeh1}
H_{\text red}=&&\frac{1}{2g^2}\pi_i^T\pi_i^T
+\frac{1}{g^2}\pi_i^T\partial_i\pi
+\frac{1}{2g^2}\partial_i\pi\partial_i\pi
\nonumber\\
&&+\frac{1}{2}g^2\partial_iA_j^T\left(\partial_iA_j^T-\partial_jA_i^T\right)
-i\bar{\psi}\gamma^i\partial_i\psi+m\bar{\psi}\psi+e_0fA_i^TJ^i
\nonumber\\
&&+\frac{1}{2}\pi_\phi\pi_\phi+\frac{1}{2}\partial_i\phi\partial_i\phi+V(\phi)
\,,
\end{eqnarray}
where $J^i\equiv \psi^\dagger\gamma^i\psi$. 
 In the Lagrangian eq. (\ref{Lag1}), we have spotted another constraint with $\chi$ as Lagrangian multiplier. Note that this term is nothing but 
the generalized current conservation equation (\ref{current1}) that reflects
the (on-shell) $U(1)$ gauge symmetry.
Therefore, we can simply remove this constraint from the Lagrangian and the reduced 
Hamiltonian gives us all correct equations of motion. One can check that 
the equations of motion given by the reduced Hamiltonian $H_{\text red}$ are 
the same as the ones given by the original Lagrangian eq. (\ref{modQED}). 
For instance, in the gauge field sector, we have 
\begin{eqnarray}
\dot{A}_i^T&=&\frac{\partial H_{\text red}}{\partial\pi_i^T}=\frac{1}{g^2}\pi_i^T+\frac{1}{g^2}\partial_i\pi,\\
\dot{\pi}_i^T&=&-\frac{\partial H_{\text red}}{\partial A_i^T}=\partial_j\left(g^2\partial_i A_j^T-g^2\partial_j A_i^T\right)-e_0fJ^i.
\end{eqnarray}
These equations of motion are the same as the ones from the original Lagrangian 
after adopting the Coulomb gauge $A_0=0$ and $\chi=0$. Similarly, we can 
also check that the equation of motion of scalar field $\phi$ and fermions are 
also consistent with the ones derived from the original Lagrangian. 
Note that the reduced Hamiltonian $H_{\text red}$ is in the quadratic form and 
thus bounded from below. We thus conclude that our theory is ghost free 
at the classical level.

 \section{Conclusion and Discussion}\label{cd}
In this paper, we considered a novel type of $U(1)$ gauge field theory, 
of which the action is invariant under local $U(1)$ gauge transformations
only when the equations of motion are imposed. We call it an on-shell
$U(1)$ gauge theory.
We applied Dirac's method of Hamiltonian analysis to clarify 
the number of degrees of freedom in this theory. 
We have spotted 1 first-class constraint and 2 second-class constraints 
in the gauge field sector, in contrast to the stand QED which has 2 first 
class constraints.  Apart from this subtle difference, 
the result is that the number of degrees of freedom in the photon sector
remains the same, that is, there are only 2 transverse degrees of freedom
despite the fact that the local $U(1)$ symmetry is broken. 
Adding the degrees of freedom of the fermion and scalar sectors,
 we conclude that  the  number of degrees of freedom in our theory 
is  the same as the usual scalar plus $U(1)$ gauge theory.
No longitudinal mode is found in the photon of our theory. 
Nor the seemingly new constraint (\ref{current3}) does not kill
the physical degrees of freedom.
We have also derived the reduced Hamiltonian
in the Coulomb gauge and found that it is bounded from below. 
Therefore our theory is free from the ghost problem.

We should mention that our analysis is classical,
and thus it is still premature to claim the validity of our
on-shell $U(1)$ gauge theory. 
It is necessary to check whether a quantum anomaly appears at loop level 
and spoils our gauge symmetry. And if so, if there is a mechanism
or a new ingredient to cancel the anomaly.
We plan to come back to this issue in our future work.

\begin{acknowledgments}
We are grateful to the participants of the workshop ``Current Themes in 
High Energy Physics and Cosmology" at the Niels Bohr Institute in August 2016,
where the issue dealt in this work was raised.
We would like to thank Slava Mukhanov for useful comments on an earlier version
of this work. CL would like to thank Shinji Mukohyama, Ryo Saito, Yi Wang for 
useful discussions. 
This work was supported in part by the MEXT KAKENHI Nos.~15H05888 and 15K21733,
and by the JSPS KAKENHI No.~15F15321.
CL is supported by the JSPS Postdoctoral Fellowship for Overseas Researchers.
\end{acknowledgments}

\end{document}